\documentclass[prb,twocolumn,showpacs,superscriptaddress]{revtex4}
\usepackage{graphicx}
\usepackage{amsmath}
\usepackage{amssymb}
\usepackage{color}
\begin{document}
\preprint{Preprint}

\title{Study of pressure effect on the magnetic penetration depth in MgB$_{2}$}

\author{D. Di Castro}
 \email[Email: ]{dicastro@physik.unizh.ch}

\affiliation{Physik-Institut der Universit\"at Z\"urich,
Winterthurerstrasse 190, CH-8057 Z\"urich, Switzerland}

\affiliation{INFM-Coherentia and Dipartimento di Fisica,
Universita' di Roma "La Sapienza", P.le A. Moro 2, I-00185 Roma,
Italy}

\author{R. Khasanov}

\affiliation{Physik-Institut der Universit\"at Z\"urich,
Winterthurerstrasse 190, CH-8057 Z\"urich, Switzerland}

\affiliation{Paul Scherrer Institute, CH-5232 Villigen PSI,
Switzerland}

\author{C. Grimaldi}
\affiliation{Ecole Polytechnique F\'ed\'erale de Lausanne, LPM,
CH-1015 Lausanne, Switzerland}

\author{J. Karpinski}

\author{S.~M. Kazakov}

\affiliation{Solid State Physics Laboratory, ETH, CH-8093
Z\"urich, Switzerland}

\author{H. Keller}

\affiliation{Physik-Institut der Universit\"at Z\"urich,
Winterthurerstrasse 190, CH-8057 Z\"urich, Switzerland}

%


%


\begin{abstract}
A  study of the pressure effect on the magnetic penetration depth
$\lambda$  in  polycrystalline MgB$_{2}$ was performed by
measuring the temperature dependence of the magnetization under an
applied pressure of 0.15 and 1.13 GPa. We found that
$\lambda^{-2}$ at  low temperature  is only slightly affected by
pressure [$\frac{\Delta \lambda^{-2}}{\lambda^{-2}} = 1.5(9)\%$],
in contrast to cuprate superconductors, where, in the same range
of pressure, a very large effect on $\lambda^{-2}$ was found.
Theoretical estimates indicate that most of the pressure effect on
$\lambda^{-2}$ in MgB$_2$ arises from the electron-phonon
interaction.
\end{abstract}



\maketitle
%
%

Shortly after the discovery of the superconductivity in MgB$_{2}$
at 39 K, several investigations of the pressure dependence of the
superconducting critical temperature $T_{c}$ were carried out.
\cite{Monteverde,Lorenz,Saito,Tomita} Indeed,  the magnitude and
sign of d$T_{c}$/d$p$ may indicate  a way  to rise $T_{c}$ at ambient
pressure, and moreover,
help to understand the superconducting pairing mechanism. So far,
all these studies show that $T_{c}$ decreases with increasing
pressure, with a rate depending on the method and the pressure
medium used. The first hydrostatic measurement of $T_{c}(p)$ up to
0.7 GPa, reveals that $T_{c}$ decreases reversibly under
hydrostatic pressure at the rate d$T_{c}$/d$p$ =  $-$ 1.11(2)
K/GPa. \cite{Tomita} The behavior of $T_{c}(p)$ with pressure in
MgB$_{2}$ was attributed to the pressure induced lattice
stiffening (increase of the phonon frequency),
\cite{Deemyad,Goncharov}
 rather than to the decrease in the electronic
density of states N(E$_{F}$), that is only moderately affected by
 pressure. \cite{Loa} Comparison with theoretical calculations
supported the view that MgB$_{2}$ is a BCS superconductor with
moderately strong electron-phonon coupling. \cite{Deemyad}

Apart from $T_{c}$, an other relevant superconducting parameter is
the magnetic field penetration depth $\lambda$. In fact the
so-called superfluid density  $\lambda^{-2}$  is related to the
Fermi velocity and to the density of charge carriers, and its
temperature dependence gives information on the symmetry and on
the magnitude of the superconducting gap. A study of  pressure
effects on $\lambda^{-2}(0)$ can give important informations on
how the electronic degrees of freedom are affected by lattice
modifications and  on the nature of the electron-phonon
coupling. Indeed, in cuprates high temperature superconductors
(HTS),   a huge pressure effect on $\lambda^{-2}(0)$ was found,
in particular in  YBa$_{2}$Cu$_{4}$O$_{8}$ (Y124).
\cite{Khasanov}  Part of this effect was attributed to the strong
renormalization of the Fermi velocity, and therefore to the effective
mass, due to the non-adiabatic coupling of the electrons to the
lattice. This result is in agreement with the substantial oxygen
isotope effect found in cuprates
\cite{Zhao95,Hofer00,Khasanov02,Khasanov03,Khasanov03b}
and can be interpreted in the framework of  non-adiabatic theory
of superconductivity. \cite{Grimaldi}

In this paper we report  measurements  of the magnetic penetration
depth under pressure in polycrystalline MgB$_{2}$. The temperature
dependence of  $\lambda^{-2}(T)$ was extracted from the Meissner
fraction $f$ measured in a low magnetic field. A small pressure
($p$) effect on $\lambda^{-2}(0)$  was found [$\frac{\Delta
\lambda^{-2}}{\lambda^{-2}} = 1.5(9)\%$], for pressure ranging
from 0.15 to 1.13 GPa. Theoretical calculations of the pressure
effect on $\lambda(0)$  confirm the smallness of this effect
[$\frac{\Delta \lambda^{-2}}{\lambda^{-2}} \simeq 1.4\%$]. These
results contrast with the huge effect ($\sim$ 40 \%) on
$\lambda^{-2}(0)$ found in the non-adiabatic cuprate
superconductor Y124 in the same range of pressure \cite{Khasanov}.

The MgB$_{2}$ powder sample was prepared by solid state reaction
in flowing argon. As starting materials we used Mg flakes
and amorphous boron (Alfa Aesar).  A pellet with starting composition Mg$_{1.1}$B$_{2}$
was placed in a  BN crucible and fired in a tube furnace under pure
Ar gas. The sample was heated for one hour at 600$^{\circ}$C, one
hour at 800$^{\circ}$C, and one hour at 900$^{\circ}$C.

The sample was first ground and then sieved in order to obtain a
small grain size $R$$<$10 $\mu$m, needed for the determination of
$\lambda$ from the Meissner fraction measurements.  The
hydrostatic pressure was produced in a copper-beryllium piston
cylinder clamp, especially designed for magnetization measurements
under pressure (see Ref [\onlinecite{Straessle}]). The sample was
put in a teflon cylinder and the pressure cell was then filled
with Fluorinert FC77 as pressure transmitting medium. The Meissner
fraction was calculated from low field magnetization measurements
(field cooling) performed with a commercial Superconducting
Quantum Interference Device. In the approximation of spherical
grains with average radius $R$, the Meissner fraction can be
related to the magnetic penetration depth via the Shoenberg
formula: \cite{Shoenberg}
\begin{equation}
  \frac{\chi}{\chi_0}= \left[1-3\left(\frac{\lambda(T)}{R}\right)\coth\left(\frac{R}{\lambda(T)}\right)+3\left(\frac{\lambda(T)}{R}\right)^{2}\right].
\label{f}
\end{equation}
 Any change in  $\chi/\chi_0$ due to  pressure can be attributed  mainly to
a change  of $\lambda(T)$, rather than to a change of  R, which is
practically pressure independent. The sample was measured at low
pressure $p_L$ = 0.15 GPa and at the highest pressure available
$p_H$ = 1.13 GPa. The pressure values were determined from a
preliminary calibration measurement, where the same sample was
measured in the pressure cell together with a small piece of lead.
The pressure was detected by measuring the $T_{c}$ shift of lead
($T_{c}$(0kbar)=7.2K). We found d$T_{c}$/d$p$ = $-$ 1.24(5) K/GPa,
in good agreement with previous results. \cite{Tomita} The real
zero-pressure measurement could not be performed in the pressure
cell, since, in order to seal the  cell, at least a small pressure
had to be applied.

\begin{figure}
\input{epsf}
\epsfxsize 8cm
\centerline{\epsfbox{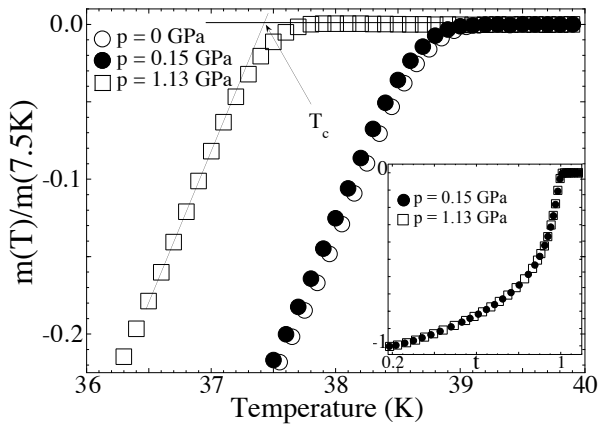}} \caption[~]{Field cooled
($0.5\,{\mathrm{mT}}$) normalized magnetization of MgB$_{2}$ as a
function of temperature in the vicinity of $T_{c}$ for $p_{Z}$
(open circles), $p_{L}$ (filled circles), and $p_{H}$ (open
squares). The inset show the full normalized magnetization data
for $p_{L}$ and $p_{H}$ as a function of the reduced temperature
$t = T/T_c$. Note that some of the data points were dropped for
clarity.} \label{fig1}
\end{figure}

In Fig. \ref{fig1} the temperature dependence of the normalized
magnetization of MgB$_{2}$ at $p_L$ and $p_H$ is shown in the
vicinity of $T_{c}$, together with the zero pressure $p_Z$
measurement performed on the same sample in a quartz tube. The
$T_{c}$'s were obtained from the intercept of the linear
extrapolations (see Fig.\ref{fig1}): $T_{c}(p_Z)$ = 38.73(3) K,
$T_{c}(p_L)$ = 38.66(3) K, $T_{c}(p_H)$ = 37.43(3) K. The
magnetization curves shift systematically with increasing pressure
towards lower temperature. In the inset of  Fig.\ref{fig1} we show
the normalized magnetization as a function of the reduced
temperature $t = T/T_c$ for $p_L$ and $p_H$. The identical
temperature dependences for both
 pressure values indicate the absence of stresses in the sample
due to pressure.

From the magnetization the Meissner fraction was extracted.
Because of the unknown mass of the sample in the pressure cell,
the Meissner fraction for the lowest pressure  ($p_L$=1.5 kbar)
was normalized at low temperature with the value obtained by
measuring the same sample in a quartz tube at zero pressure
$p_{Z}$. The values of $\lambda^{-2}$ were then calculated using
the Shoenberg formula (Eq. (\ref{f})). The $p_Z$ and $p_L$ data
were normalized at low temperature to the absolute value of
$\lambda^{-2}$ measured with the muon spin rotation technique
($\mu$SR) on the same sample at zero pressure. \cite{DiCastro2}

We define the shift of $\lambda^{-2}$ between two different
pressures $p_{L}$ (lower) and $p_{H}$ (higher) at a temperature
$T$ as:
\begin{equation}
\frac{\Delta \lambda^{-2}}{\lambda^{-2}}\equiv
\frac{\lambda^{-2}(p_{H},T)-\lambda^{-2}(p_{L},T)}{\lambda^{-2}(p_{L},T)}
\label{Delta}
\end{equation}

\begin{figure}[htb]
\includegraphics[width=1.0\linewidth]{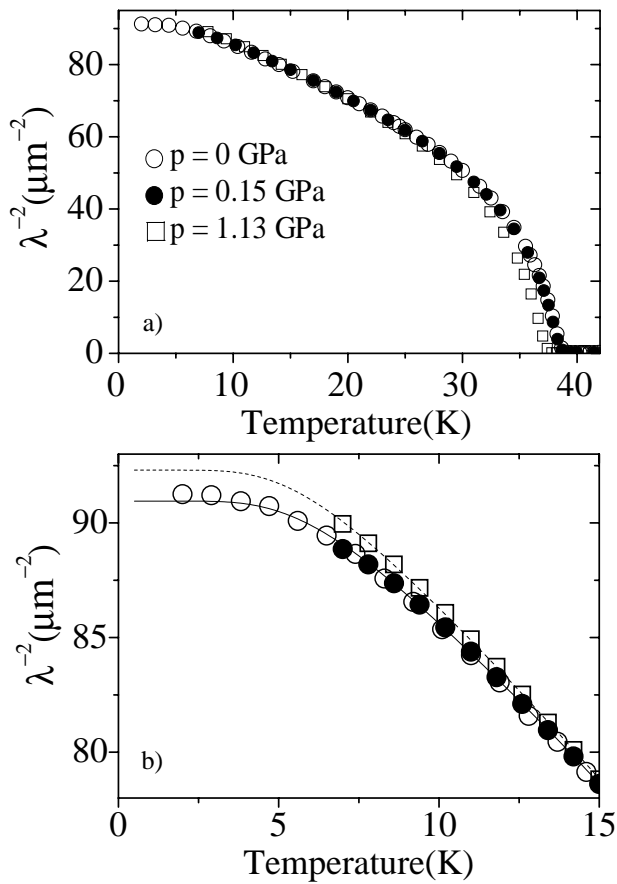}
\caption[~]{a) $\lambda^{-2}$ as a function of temperature for
$p_{Z}$ = 0 GPa (open circles), $p_{L}$ = 0.15 GPa (filled
circles), and $p_{H}$ = 1.13 GPa (open squares). b) Low
temperature region on a larger scale. The error bars (not shown)
for temperatures below 10 K,  are about two times the symbol size
and decrease with increasing temperature. The solid and dashed
lines are fits to the $p_{L}$ and $p_{H}$ data, respectively,
using Eq. \ref{twogapmodel}. Note that the data points are much
more dense than what is shown in this figure, since part of them
were dropped for clarity.}
 \label{lambda}
\end{figure}

The temperature dependences of $\lambda^{-2}$ for $p_{L}$ and
$p_{H}$, are shown in Fig. \ref{lambda}, together with the $p_{Z}$
data measured in the quartz tube. The $\lambda^{-2}(T)$ curves for
$p_{Z}$ and $p_{L}$  overlap perfectly for temperatures not too
close to $T_{c}$, where a small pressure shift is present.
Unfortunately, we were not able to obtain reliable data below 7K
because of a large background signal due to lead impurities in the
pressure cell. Therefore, while for the $p_Z$ measurement in the
quartz tube we show the data down to 2 K, for the $p_L$ and the
$p_H$ measurements we show the data only down to 7 K (see lower
panel of Fig. \ref{lambda}). Due to the pressure effect on
$T_{c}$, the curves for $p_{L}$ and $p_{H}$ are clearly distinct
at high temperature, but they merge at low temperature. The error
bars on $\lambda^{-2}$ (not shown) were determined considering
that magnetization measurements are reproducible within $\approx$
0.5 $\%$.

The shift of $\lambda^{-2}$ at the lowest
temperature available $T \simeq$ 7 K between $p_{L}$ = 0.15
GPa and $p_{H}$ = 1.13 GPa (see Eq. \ref{Delta}) is:
$\Delta \lambda^{-2}/\lambda^{-2} = 1.2(1.1)\%$.
This result shows that only a very small pressure effect on
$\lambda^{-2}$ is present. The pressure results are  summarized in
Table \ref{summary}.

To determine the pressure effect on the zero temperature
$\lambda^{-2}(0)$, a fit to the experimental data is needed. It
was well established that MgB$_{2}$ is a two band superconductor
with two superconducting gaps of different size, the larger one
originating from  the 2D $\sigma$-band and the smaller one from
the 3D $\pi$-band \cite{Gonnelli,Bouquet01,Szabo01,Souma03}.
Taking this into account, the temperature dependence of $\lambda$,
in the low temperature range, can be written in the form
\cite{Kim02}:

\begin{eqnarray}
\label{twogapmodel} \lambda^{-2}(T)&=&\lambda^{-2}(0)\left[1- w
\left( \frac{2\pi\Delta_{1}(0)}{k_{B} T}\right)^{1/2}\!\!
\exp\!\left(-\frac{\Delta_{1}(0)}{k_{B}T}\right)\right. \nonumber \\
&-& \left.(1-w) \left( \frac{2\pi\Delta_{2}(0)}{k_{B}
T}\right)^{1/2}\!\! \exp\!\left(-\frac{\Delta_{2}(0)}{k_{B}
T}\right)\right]
\end{eqnarray}
Here, $\Delta_{1}$ and $\Delta_{2}$ are the zero temperature small
and large gap associated to the $\pi$ and $\sigma$ bands,
respectively, and $w=\lambda_1^{-2}(0)/\lambda^{-2}(0)$.

The zero pressure  data were then fitted to  Eq. \ref{twogapmodel}
up to 22 K. The fit gave the following results: $\lambda^{-2}(0)$
= 91.1(2) $\mu$ m$^{-2}$, $w$ = 0.22(3); $\Delta_{1}$ = 2.3(2)
meV, $\Delta_{2}$ = 7.4(4) meV. The estimated gaps
 are in good
agreement with previous  results from penetration depth
measurements \cite{Kim02,Manzano}.

Because of the lack of  data below 7 K for $p_{L}$  and $p_{H}$,
these data were analysed with fixed $w$ obtained from the  fit to
the zero-pressure data. Here it is assumed that $w$  is not
affected by pressure up to 1.13 GPa. This assumption will be shown
below to be correct  by our model calculations [see
Eq.(\ref{th5})]. The other fitting parameters ($\lambda^{-2}(0)$
and the gaps) were left free. As shown by the solid and dashed
lines in  panel b) of Fig. \ref{lambda}, the data are  well
described by Eq.\ (\ref{twogapmodel}). In particular the fitted
curve of  the $p_{L}$ data (solid line) follows very well the
$p_{Z}$ data below 7 K, as expected. The fit yields for
$\lambda^{-2}(0)$: $\lambda^{-2}(0)_{p_{L}} = 90.9(5) \mu m^{-2}$
and $\lambda^{-2}(0)_{p_{H}} = 92.3(6) \mu m^{-2}$ (see Table
\ref{tablesummary}). So that the relative shift of $\lambda^{-2}$
at $T = 0$ between  $p_{L}$ = 1.5 kbar and $p_{H}$ = 11.3 kbar is
[see Eq. (\ref{Delta})]:
\begin{equation}
\frac{\Delta \lambda^{-2}}{\lambda^{-2}} = 1.5(9)\%
\label{Deltalambda}
\end{equation}
This result indicates that there is a small, but not zero,
pressure effect on the magnetic penetration depth at 0 K in the
range $p$ = 0.15 - 1.13 GPa.
A summary of the pressure results are  reported in Table \ref{summary}.
\begin{table}
\caption[~]{Summary of the experimental and theoretical
estimations of the pressure shift of $\lambda^{-2}$.} \label{summary}
\begin{ruledtabular}
\begin{tabular}{c|c|c|c|c} 
%
&$T$&
$p_{L}$&$p_{H}$&$\frac{\lambda^{-2}(p_{H},T)-\lambda^{-2}(p_{L},T)}{\lambda^{-2}(p_{L},T)}$\\
&(K)&(GPa)&(GPa)&\%\\
\hline exper.&$\simeq$7&0.15&1.13&1.2(1.1)
\\
fit&0&0.15&1.13&1.5(9)
\\
theory&0&0.15&1.13&$\simeq 1.4\div 1.7$
\end{tabular}
\end{ruledtabular}
\end{table}
All the  fitting parameters are
summarized in the Table \ref{tablesummary}. The two gaps show a
small decrease with increasing pressure,  in agreement with the
corresponding decrease of  $T_{c}$. In Table \ref{tablesummary} we
added also the ratio 2$\Delta/k_BT_c$ calculated for $\Delta_1$
and $\Delta_2$ at each pressure. This ratio results to be pressure
independent within errors.
\begin{table*}
\caption[~]{Fitting parameters obtained from the fit of the
experimental data showed in Fig. \ref{lambda} by the Eq.
(\ref{twogapmodel}).} \label{tablesummary}
\begin{ruledtabular}
\begin{tabular}{c|c|c|c|c|c|c} 
%
$p$&$\lambda^{-2}(0)$&
$\Delta_{1}(0)$&$\Delta_{2}(0)$&$w$&2$\Delta_{1}(0)/k_{B}T_c$&2$\Delta_{2}(0)/k_{B}T_c$\\
GPa&($\mu$m$^{-2}$)&(meV)&(meV)&\\
\hline  0 &91.1(2)&2.3(2)&7.4(4)&
0.22(3)&1.38(12)&4.43(24)\\
 0.15 &90.9(5)&2.4(1)&7.3(1)&&1.44(6)&4.38(6)
\\
 1.13 &92.3(6)&2.3(1)&7.0(1)&&1.43(6)&4.34(6)
\\
\end{tabular}
\end{ruledtabular}
\end{table*}

Considering that $p_H-p_L \simeq 1$ GPa,  Eq. (\ref{Deltalambda})
indicates that $d\ln\lambda^{-2}(0)/dp$ is of the order of few
$\%/$GPa. Since $\lambda^{-2}(0)$ is proportional to the squared
plasma frequency, $\omega_p^2$, then a free electron gas estimate
would give $d\ln\lambda^{-2}(0)/dp=1/B\simeq 0.65\%/$GPa, where we
have used $\omega_p^2\propto 1/\Omega$, being
$B=-dp/d\ln\Omega\simeq 155$ GPa  the bulk modulus of
MgB$_2$,\cite{Goncharov} and $\Omega$  the volume of the unit
cell. As we show below, an improved estimate which takes into
account MgB$_2$ band structure effects does not change much the
free electron gas result. Let us consider the zero temperature
expression of the penetration depth:\cite{Carrington}
\begin{equation}
\label{th1} \lambda^{-2}(0)=\frac{e^2}{3\hbar \pi^2 c^2}\sum_n
\oint_{S_F^n} ds |{\bf v}_n(s)|,
\end{equation}
where the integral runs over the Fermi surface $S_F^n$ of the
$n$-th band ($n=\sigma,\pi$) and ${\bf v}_n(s)$ is the
corresponding surface bare electron velocity vector. Let us use
the model of Ref. \onlinecite{dahm} where the $\pi$ bands are
modeled by a half-torus Fermi surface of area $S_1$ and Fermi
velocity $v_1$, while the $\sigma$ bands are approximated by a
cylindrical Fermi surface of area $S_2$ and Fermi velocity $v_2$.
Equation (\ref{th1}) then reduces simply to
$\lambda^{-2}(0)=\lambda_1^{-2}(0)+\lambda_2^{-2}(0)$, where
$\lambda_i^{-2}(0)\propto S_i v_i$, $i=1,2$, and
\begin{equation}
\label{th2}
\frac{d\ln\lambda^{-2}(0)}{dp}=w\frac{d\ln\lambda_1^{-2}(0)}{dp}+
(1-w)\frac{d\ln\lambda_2^{-2}(0)}{dp},
\end{equation}
where $w=\lambda_1^{-2}(0)/\lambda^{-2}(0)=S_1v_1/(S_1v_1+S_2v_2)$
is the parameter introduced in Eq.(\ref{twogapmodel}). Within the
same approximations, the electron density of states at the Fermi
level reads $N_F=N_1+N_2$, where $N_i\propto \Omega S_i/v_i$ is
the partial density of states for the band $i=1,2$. Therefore,
since $\lambda^{-2}_i(0)\propto S_i v_i\propto \Omega S_i^2/N_i$,
and considering that $S_1$ and $S_2$ scale as $\Omega^{-2/3}$,
Eq.(\ref{th2}) reduces to:
\begin{equation}
\label{th4} \frac{d\ln\lambda^{-2}(0)}{dp}\!=
\!\frac{1}{3B}\!-\!\left[\frac{1-w+\eta
w}{\eta+(1-\eta)N_2/N_F}\right]\!\frac{d\ln N_F}{dp},
\end{equation}
where we have introduced the parameter $\eta=(d\ln N_1/dp)/(d\ln
N_2/dp)$ whose calculated value ranges between $\eta\simeq
1$,\cite{medve} and $\eta\simeq 0$.\cite{vogtraz} Hence, by
setting $d\ln N_F/dp\simeq -0.31\%/$GPa, \cite{Loa} $w=0.22$, and
$N_2/N_F\simeq 0.4$,\cite{liu} from Eq.(\ref{th4}) we obtain
$d\ln\lambda^{-2}(0)/dp \simeq 0.8\%/$GPa or $0.5\%/$GPa according
to whether $\eta=0$ or $\eta=1$, respectively. By the same token,
it is easy to show that:
\begin{equation}
\label{th5}
\frac{dw}{dp}=\frac{w(1-w)(1-\eta)}{\eta+(1-\eta)N_2/N_F}\frac{d\ln
N_F}{dp}\simeq (0\div -0.1)\%/{\rm GPa},
\end{equation}
confirming the assumption $dw/dp=0$ used in obtaining
Eq.(\ref{Deltalambda}).

Our estimate $\Delta\lambda^{-2}/\lambda^{-2}\simeq (0.5\div
0.8)\%$ does not deviate from the free eletron gas result,
suggesting that other factors than band structure should be
considered in order to explain the measured value reported in
Eq.(\ref{Deltalambda}). It is then natural to consider the
electron-phonon interaction $\lambda_{el-ph}$. The  pressure
effect on $\lambda_{el-ph}$ in MgB$_2$ is mainly due to a pressure
induced hardening of the optical phonon modes. The electron-phonon
renormalized penetration depth is
$\lambda^{*-2}(0)=\lambda^{-2}(0)/(1+\lambda_{el-ph})$, where
$\lambda^{-2}(0)$ is the bare quantity we have considered before.
Hence:
\begin{equation}
\label{th6} \frac{d\ln\lambda^{*-2}(0)}{dp}\simeq (0.5\div
0.8)\%/{\rm
GPa}-\frac{\lambda_{e-ph}}{1+\lambda_{e-ph}}\frac{d\ln\lambda_{e-ph}}{dp},
\end{equation}
which, by using $d\ln\lambda_{e-ph}/dp\simeq -1.7\%/$GPa,
\cite{Loa} and $\lambda_{el-ph}\simeq 1$,\cite{liu} corresponds to
$\Delta\lambda^{-2}/\lambda^{-2}\simeq (1.4\div 1.7)\%$, in better
agreement with Eq.(\ref{Deltalambda}). This simple analysis
evidences therefore that the electron-phonon interaction provides
the main contribution to the pressure effect on the zero
temperature penetration depth. This theoretical estimate of the
pressure shift of  $\lambda(0)^{-2}$ is reported in the Table
\ref{summary} for a direct comparison with the experimental
finding.

Similar experiment on the Y124, in the same range of
pressure, gives a very large effect on $\lambda(0)^{-2}$ ($\sim
40 \%$).\cite{Khasanov} Arguments are used there to deduce that part
($\sim 30 \%$) of this large  effect is due to the pressure
dependence of the Fermi velocity, because of the non-adiabatic
electron-lattice coupling. From the results of the present work,
we can therefore argue that such non-adiabatic effects on $\lambda(0)^{-2}$ are
negligible in MgB$_{2}$, as previously demonstrated by $\mu$SR
measurements.\cite{DiCastro}

In summary, we studied the  pressure effect on the magnetic
penetration depth at low temperature in polycrystalline MgB$_{2}$.
We found that pressure up to 1.13 GPa induces a small positive
change in $\lambda(0)^{-2}$ [$\frac{\Delta
\lambda^{-2}}{\lambda^{-2}} = 1.5(9)\%$], which is suggested to be
due mostly to a pressure change of the electron-phonon coupling.

The authors are grateful to S. Kohout for the help during the
measurements, T. Schneider for  very fruitful discussions, R. Brutsch and D. Gavillet for the measurements
of the grain size distribution of the powder sample. This work was
supported by the Swiss National Science Foundation and partly by the NCCR
Program MaNEP sponsored by the Swiss National Science Foundation.

\newcommand{\noopsort}[1]{} \newcommand{\printfirst}[2]{#1}
  \newcommand{\singleletter}[1]{#1} \newcommand{\switchargs}[2]{#2#1}

\end{document}